\documentclass[runningheads]{llncs}
\usepackage[T1]{fontenc}
\usepackage{graphicx}
\usepackage[most]{tcolorbox}
\usepackage{multirow}
\usepackage{booktabs}
\usepackage{makecell}
\usepackage{tabularx}  

\begin{document}
\title{Dataset Ownership in the Era of Large Language Models}

\author{Kun Li \inst{1}
\and
Cheng Wang \inst{1}
\and
Minghui Xu \inst{1*}
\and
Yue Zhang \inst{1}
\and
Xiuzhen Cheng \inst{1}
}
\authorrunning{K. Li et al.}
%
\institute{Shandong University, Jinan, China\\
\email{kunli@sdu.edu.cn, 1933831694@qq.com, \{mhxu, zyueinfosec, xzcheng\}@sdu.edu.cn}\\
* Corresponding author: Minghui Xu
}
\maketitle              
\begin{abstract}
As datasets become critical assets in modern machine learning systems, ensuring robust copyright protection has emerged as an urgent challenge. Traditional legal mechanisms often fail to address the technical complexities of digital data replication and unauthorized use, particularly in opaque or decentralized environments. This survey provides a comprehensive review of technical approaches for dataset copyright protection, systematically categorizing them into three main classes: non-intrusive methods, which detect unauthorized use without modifying data; minimally-intrusive methods, which embed lightweight, reversible changes to enable ownership verification; and maximally-intrusive methods, which apply aggressive data alterations, such as reversible adversarial examples, to enforce usage restrictions. We synthesize key techniques, analyze their strengths and limitations, and highlight open research challenges. This work offers an organized perspective on the current landscape and suggests future directions for developing unified, scalable, and ethically sound solutions to protect datasets in increasingly complex machine learning ecosystems.

\keywords{Dataset Ownership \and Watermarking \and Copyright Protection \and Large Language Models (LLMs).}
\end{abstract}

\section{Introduction}
In the age of machine learning (ML) and big data, datasets have become invaluable assets that power advancements across scientific, industrial, and commercial domains. However, dataset protection has become a significant challenge, as traditional methods of intellectual property (IP) protection are ill-equipped to address the complexities of modern machine learning. Unlike physical assets, datasets are inherently replicable, easily shared, and often used across various machine learning models, raising concerns about their ownership and misuse.

Traditional methods of data protection, such as copyright laws and digital rights management (DRM), have proven insufficient in safeguarding datasets. These methods are typically designed for tangible assets and fail to address the unique characteristics of digital datasets used in machine learning systems. While legal and DRM solutions offer some level of protection, they do not address the specific technical issues involved in verifying dataset ownership, especially in decentralized and distributed environments where data is easily replicated and reused.

The rise of large language models (LLMs) has intensified focus on data security, with most existing research centered on privacy risks (e.g., differential privacy, secure multi-party computation) and model protection (e.g., watermarking, federated learning) \cite{YAN2025100300,YAO2024100211}. However, the issue of data copyright and ownership verification, crucial for ensuring the legitimacy of training dataset usage, has received relatively less attention. Unlike model protection—where techniques primarily safeguard weights or outputs—the security of training datasets demands novel approaches to address challenges like ownership attribution and unauthorized reuse.

In this survey, we focus on dataset copyright protection within the context of modern machine learning systems, providing a comprehensive review of the technical methods used to safeguard dataset ownership. Specifically, we categorize these methods into three types: non-intrusive, minimally-intrusive, and maximally-intrusive, analyzing each in terms of its effectiveness, robustness, and applicability across various types of data. By doing so, we aim to provide a clear framework for understanding the strengths and limitations of current dataset protection techniques and to offer insights into their practical use in machine learning environments.

The primary contributions of this survey are:
\begin{enumerate}
    \item A detailed classification of dataset protection techniques into three categories, offering a structured approach to understanding the strengths, limitations, and trade-offs of each method.
    \item An in-depth analysis of the techniques within each category, highlighting their relevance, robustness, and suitability for different use cases, particularly in the context of machine learning.
    \item A focus on emerging challenges in the field, especially as they relate to the protection of datasets used in large-scale machine learning models and foundation models.
\end{enumerate}

Moreover, our survey highlights several key insights and areas for future research. Notably, we observe that most existing work has concentrated on image datasets, leaving text, audio, and other data types underexplored. This gap presents an exciting opportunity for cross-domain protection techniques, ensuring dataset ownership across diverse forms of data used in machine learning. Additionally, the rise of large-scale models calls for specialized protection methods that address the unique challenges posed by these systems. Ethical and legal concerns surrounding data privacy and ownership transparency are often overlooked, but integrating these factors into future research will be crucial to ensuring responsible dataset protection. Lastly, scalability and efficiency remain significant challenges in applying protection methods to large-scale datasets, necessitating the development of efficient, scalable techniques that do not compromise model performance.

Through this survey, we aim to contribute to the development of more robust, scalable, and ethically responsible solutions for dataset protection in the rapidly evolving field of machine learning. The remainder of this survey is organized as follows: Section \ref{sec:related} reviews the existing literature on dataset protection techniques, including auditing, watermarking, and related approaches. Section \ref{sec:1} discusses non-intrusive methods that verify ownership without altering the data. Section \ref{sec:2} covers minimally-intrusive methods, which involve lightweight modifications for ownership verification while preserving data integrity. Section \ref{sec:3} explores maximally-intrusive methods, which apply aggressive alterations to enforce stronger control over dataset usage. Section \ref{sec:challenge} highlights key challenges and future research directions. Finally, Section \ref{sec:conclusion} concludes the survey, summarizing the insights and proposing avenues for further work in dataset copyright protection. 

\section{Related Work}\label{sec:related}
The protection of dataset ownership has become a critical research area, particularly in the context of the rapid development of machine learning and big data applications. Existing literature offers valuable insights into various aspects of dataset protection. For instance, Du \textit{et al}.\cite{sok1} provides an overview of dataset auditing techniques, categorizing them into intrusive and non-intrusive methods. While the paper thoroughly explores the advantages and limitations of these approaches, it primarily focuses on the auditing and verification of dataset usage. However, it does not delve deeply into the direct protection of dataset ownership through technologies such as watermarking or encryption. Similarly, Chandrasekaran \textit{et al}. \cite{sok2} examines the issue of data ownership from a machine learning governance perspective, highlighting the importance of accountability and assurance measures. Yet, the discussion lacks a detailed exploration of specific technical solutions for protecting dataset ownership.

In the realm of image data protection, Ray \textit{et al}. \cite{sok3} extensively reviews watermarking techniques, particularly focusing on their applications in copyright protection. However, its scope is largely confined to image data, with less emphasis on the protection of other types of datasets used in machine learning. In the context of security and privacy, Papernot \textit{et al}.\cite{sok4} introduces a comprehensive threat model and analyzes defense mechanisms. While the paper provides important insights into model-level security, it does not specifically address the issue of dataset copyright protection. Lastly, Asswad \textit{et al}.\cite{sok5} explores data ownership from legal, ethical, and technical perspectives. While the paper offers a broad conceptual discussion on data protection, it does not focus on the technical mechanisms necessary for safeguarding dataset ownership.

Although these studies provide valuable contributions to the field of dataset protection, they primarily focus on auditing, model security, or specific protective methods, often overlooking a comprehensive approach to dataset protection across diverse data types and machine learning contexts. In contrast, this work presents a more holistic framework for dataset protection, categorizing methods into non-intrusive, minimally-intrusive, and maximally-intrusive techniques, and providing a detailed analysis of each. Our work not only addresses the gap in existing research regarding dataset protection in the context of large-scale models but also integrates discussions on the ethical, legal, and technical challenges associated with dataset ownership. This comprehensive approach advances the field by offering a more nuanced perspective on the protection of datasets in complex machine learning systems.

\section{Overview}\label{sec:overview}
The task of protecting dataset ownership in machine learning poses unique challenges distinct from those faced in traditional digital content protection. Datasets are not merely static artifacts; they serve as the foundation for model training, adaptation, and knowledge extraction, making their unauthorized use difficult to detect and even harder to control. Recent years have seen a proliferation of technical methods designed to safeguard datasets against misuse, but the field remains fragmented, with diverse approaches targeting different stages, modalities, and threat models.

This survey is particularly motivated by the rise of large-scale foundation models, which have fundamentally reshaped how data is leveraged in machine learning. Traditional dataset protection methods, many of which focus on perceptual or sample-level integrity, often fall short in the face of large models’ capacity to absorb and generalize from massive training corpora, making it increasingly challenging to trace or constrain the influence of specific datasets. While recent research on large model copyright protection has attracted significant attention, much of it focuses on protecting model outputs, weights, or fine-tuned variants, leaving the question of dataset-level ownership comparatively underexplored.

To address this gap, we provide a structured and comprehensive synthesis of existing dataset protection methods, categorizing them into three main paradigms based on the degree and nature of data modification involved. Specifically, we distinguish between: 
\begin{enumerate}
    \item Non-intrusive methods, which verify ownership without altering the data;
    \item Minimally-intrusive methods, which embed lightweight, often reversible signals;
    \item Maximally-intrusive methods, which introduce substantial, often adversarial modifications to proactively block unauthorized use
\end{enumerate}

This tripartite division reflects not only the technical mechanisms employed but also the conceptual stance each approach takes toward balancing usability, robustness, and enforcement.

Table \ref{tab:dataset-protection-comparison} summarizes their key features, including modification level, and typical examples. Our survey builds on a comprehensive review of representative works drawn from leading venues in machine learning, computer vision, multimedia, and security. Rather than aiming for exhaustiveness, we focus on methods that exemplify distinct approaches or introduce influential techniques, ensuring that the synthesis captures both the breadth and depth of the field. In doing so, we aim not only to clarify the current landscape but also to highlight how emerging challenges in the era of large models demand new thinking about the boundaries, limitations, and future directions of dataset protection research.

\begin{table}[ht]
\centering
\caption{Comparison of Dataset Protection Approaches}  
\label{tab:dataset-protection-comparison}  
\resizebox{\textwidth}{!}{%
\begin{tabular}{@{}l l l l@{}}
\toprule
\textbf{Category} & \textbf{Modification \newline Level} & \textbf{Example Techniques} & \textbf{Representative \newline References} \\ \midrule
\multirow{3}{*}{Non-intrusive} & \multirow{3}{*}{None} & Zero-watermarking & \cite{ren2023zero,wang2019ternary}\\ \cmidrule{3-4} 
 &  & Decision boundary-based detection & \cite{DatasetInferenceforSelf-SupervisedModels,DatasetInference:OwnershipResolutioninMachineLearning}  \\ \cmidrule{3-4} 
 &  & Predictive behavior analysis & \cite{hisamoto2020membership,9724248,10175589} \\ \midrule
\multirow{3}{*}{Minimally-intrusive} & \multirow{3}{*}{Lightweight} & Reversible data hiding & \cite{cao2015high,9628041,ke2020fully,li2013novel,luo2023reversible,wang2024reversible,10.1016/j.jvcir.2024.104239,zhang2013recursive}  \\ \cmidrule{3-4} 
 &  & Clean-label watermarking & \cite{guo2023domain,li2022untargeted,li2020opensourceddatasetprotectionbackdoor,li2023black,shang2024tracking,sun2023codemark,tang2023did,wei2024pointncbw} \\ \cmidrule{3-4} 
 &  & Radioactive data & \cite{atli2022effectiveness,sablayrolles2020radioactive} \\ \midrule
\multirow{3}{*}{Maximally-intrusive} & \multirow{3}{*}{Aggressive} & Reversible adversarial examples & \cite{cao2024reversible,carlini2017towards,chen2024reversible,fang2023improving,liu2023unauthorized,szegedy2014intriguingpropertiesneuralnetworks,xue2023dataset,yin2023reversible,yin2019reversible} \\ \cmidrule{3-4} 
 &  & Black-box RAE & \cite{gao2021push,huang2024black,xiong2023black}
 \\ \cmidrule{3-4} 
 &  & Visible watermark perturbations & \cite{chen2024rae,shan2023glaze,zhao2022guided}
 \\ \bottomrule
\end{tabular}
}
\end{table}

\section{Non-Intrusive Protection}\label{sec:1}
Non-intrusive copyright protection secures dataset ownership without altering the original data. Unlike watermarking or adversarial modification methods, these approaches neither embed external signals nor perturb data but instead leverage indirect evidence or inherent data properties to verify unauthorized use. This makes them particularly appealing in scenarios where preserving data integrity is critical or where embedding signals is technically infeasible. Existing non-intrusive protection techniques can be broadly classified into two main categories: zero-watermarking and model-based detection methods.

\subsection{Zero-watermarking technology}
Zero-watermarking protects digital content by generating unique identifiers from intrinsic data features (e.g., geometric shapes, topological relationships, or attributes) without altering the original data. For example: Wang et al. \cite{wang2019ternary} developed a robust algorithm for stereoscopic images using ternary radial harmonic Fourier moments (TRHFM). Their method computes TRHFM, randomly selects components via a Logistic map to create a binary feature image, and encrypts it with a logo to generate the zero-watermark. Similarly, Ren et al. \cite{ren2023zero} designed a scheme for vector geospatial data by leveraging spatial topological relationships, combining topological and metric parameters to produce zero-watermark information.

\subsection{Model-Based Detection Methods}

\subsubsection{Decision Boundary-Based Methods}
The core objective of dataset or model theft is to steal proprietary knowledge. Maini et al. \cite{DatasetInference:OwnershipResolutioninMachineLearning} proposed leveraging original training data to detect theft: a pirated model exhibits similar prediction margins (distances to the decision boundary) on this data as the legitimate model, enabling ownership verification via boundary similarity.
However, this fails for self-supervised models lacking defined boundaries. To address this, Dziedzic et al. \cite{DatasetInferenceforSelf-SupervisedModels} developed an embedding-based method: it uses Gaussian Mixture Models to compare the log-likelihoods of training versus test data in the output embedding space, identifying theft through distribution differences.

\subsubsection{Predictive Behavior-Based Methods}
A machine learning model’s predictive behavior inherently reflects the influence of its training data, leading models trained on the same dataset to exhibit similar outputs. Liu \textit{et al}.\cite{9724248} proposed a data intellectual property protection scheme leveraging membership inference attacks (MIA). By training multiple reference models to produce universal membership fingerprints and constructing an authentication model to test predictive consistency on fingerprint samples, they detect whether a model was trained on protected data. Hisamoto \textit{et al}.\cite{hisamoto2020membership} extended MIA techniques to sequence-to-sequence models, aiming to identify whether specific data points were used to train machine translation (MT) systems; this study represents a foundational benchmark for MIA in sequence generation tasks. Additionally, Tian \textit{et al}.\cite{10175589} introduced a knowledge representation framework based on adversarial samples, where the decision boundary serves as a proxy for learned knowledge. They developed a geometric consistency method, MinAD, to generate boundary-support samples as unique knowledge representations.

\begin{tcolorbox}[title=Summary and Supplementary Note, colback=gray!10, colframe=black]
Non-intrusive copyright protection primarily employs two strategies: zero-watermarking, which exploits intrinsic data features to generate unique identifiers without altering content, and model-based detection, which analyzes predictive behavior or decision boundaries to identify unauthorized use. Blockchain technology has also emerged as a powerful solution, leveraging its decentralized, tamper-resistant, and cryptographic nature to enhance data integrity, privacy, and trust. Widely adopted across finance, healthcare, supply chains, and judicial evidence, blockchain underpins innovations like Gupta et al.'s TrailChain \cite{GUPTA2024103389}, which enhances data provenance and ownership verification \cite{9705115,YANG2024100199}.
\end{tcolorbox}

\section{Minimally-Intrusive Protection}\label{sec:2}

Minimally-intrusive protection techniques embed ownership signals into datasets through lightweight modifications, aiming to preserve data quality and model performance. Unlike non-intrusive methods, these approaches intentionally adjust the data, but in a manner designed to minimize downstream effects, enabling verifiable ownership through methods such as reversible data hiding and dataset watermarking.

\subsection{Reversible Data Hiding}
Reversible data hiding (RDH) \cite{luo2023reversible,cao2015high,zhang2013recursive,li2013novel,wang2024reversible,ke2020fully} enables embedding secret information within a carrier while ensuring the original data can be perfectly restored. This allows dataset owners to insert copyright-related information, such as author identifiers or timestamps, into multimedia content without perceptible degradation. RDH methods prioritize minimizing visual or auditory distortion. For example, Wang \textit{et al}.\cite{10.1016/j.jvcir.2024.104239} proposed an RDH scheme for color images based on prediction error value ordering (PEVO) and adaptive embedding. By applying an improved particle swarm optimization (IPSO) algorithm to generate adaptive two-dimensional mappings and optimizing these mappings using position information, their method enhances both embedding capacity and visual fidelity. Similarly, Hu and Xiang \cite{9628041} developed an RDH approach leveraging convolutional neural network (CNN) prediction and adaptive embedding. They improved prediction accuracy through novel image segmentation and a CNN-based predictor, while optimizing the embedding process via prediction error ordering (PEO) and adaptive two-dimensional mapping, thereby reducing distortion during data hiding.

\subsection{Dataset Watermarking}
Dataset watermarking protects the intellectual property of machine learning datasets by modifying selected training samples or labels so that the resulting models exhibit verifiable behaviors under specific inputs.
\subsubsection{Clean-label Watermarking}
Clean-label watermarking refers to watermarking techniques that modify the data in a way that does not directly alter the labels or minimizes such alterations, allowing the model to learn ownership signals while preserving the integrity of the dataset. This category includes Backdoor Watermarking and Domain Watermarking, both of which aim to protect datasets by embedding watermarks in a subtle and covert manner, without introducing significant errors in model predictions.

Backdoor watermarking involves embedding covert watermark samples into the dataset through poisoning attacks. These samples, when encountered by the trained model, trigger specific behaviors or outputs that can be used to verify ownership. Li \textit{et al}.\cite{li2020opensourceddatasetprotectionbackdoor,li2023black} introduced a backdoor-embedded dataset watermarking (BEDW) approach, which embeds covert watermark samples into datasets through poisoning attacks and verifies dataset usage by applying hypothesis testing on the target model. While effective, such methods introduce security risks, as adversaries could exploit hidden backdoors to manipulate model predictions. To mitigate this, Li \textit{et al}.\cite{li2022untargeted} proposed an unpredictable backdoor watermarking (UBW) scheme, where model behavior upon encountering a trigger is intentionally unpredictable, reducing the risks associated with targeted backdoor attacks.

Despite their effectiveness, backdoor watermarking techniques often require injecting mislabeled data, which can degrade model performance and may be detectable through anomaly detection. Furthermore, most approaches have been limited to image classification tasks. Addressing these limitations, Tang \textit{et al}.\cite{tang2023did} proposed a clean-label backdoor watermarking framework, which inserts a small number of covertly perturbed watermark samples. This allows models to implicitly learn a secret backdoor function, enabling ownership tracking across third-party models trained on protected datasets. Their framework has demonstrated applicability across text, image, and audio domains.

Backdoor watermarking has also been extended to specialized datasets. Wei \textit{et al}.\cite{wei2024pointncbw} proposed the PointNCBW method for point cloud datasets, perturbing non-target category samples at both the shape and point levels to embed negative triggers, thus enabling ownership verification. Shang \textit{et al}.\cite{shang2024tracking} developed a traceable watermarking scheme for machine learning cloud services, embedding encodable triggers via a clean-label backdoor framework to determine whether third-party models were trained on protected data. For code datasets, Sun \textit{et al}.\cite{sun2023codemark} introduced CodeMark, an imperceptible watermarking method designed to prevent unauthorized use by neural code completion models.

While backdoor-based watermarking can effectively verify dataset usage, it inherently introduces erroneous classification behaviors. To address this, Guo \textit{et al}.\cite{guo2023domain} proposed domain watermarking, a non-backdoor-based approach that identifies a hard-to-generalize domain as the trigger. The watermark model is trained to correctly classify these domain-specific samples, which benign models typically misclassify. To ensure stealth, the domain is carefully constructed through bi-level optimization, generating visually indistinguishable clean-label variant data that functions equivalently to domain watermark samples. Compared to backdoor watermarks, domain watermarking offers a harmless and covert alternative.

\subsubsection{Radioactive Data}
An alternative watermarking approach, radioactive data injects identifiable signals directly into the feature space of samples, rather than modifying pixel values or labels or relying on backdoor triggers. While fundamentally a form of artificial watermarking, it operates through subtle, statistically detectable modifications that persist throughout training, enabling post hoc verification without requiring explicit trigger conditions. Sablayrolles \textit{et al}.\cite{sablayrolles2020radioactive} first proposed this concept, embedding watermarks by altering the feature distributions of samples. These modifications are minimal, preserve true labels, and exert negligible impact on model performance, yet they remain detectable through statistical tests after training. Subsequently, Atli \textit{et al}.\cite{atli2022effectiveness} evaluated the effectiveness of radioactive data, finding strong performance in black-box verification but identifying limitations in white-box scenarios, particularly when the number of classes or samples per class is small.

\begin{tcolorbox}[title=Summary and Insight, colback=gray!10, colframe=black]
Minimally-intrusive protection balances ownership verification with data integrity. Reversible data hiding suits perceptual media but is less applicable to machine learning. Dataset watermarking methods—backdoor, domain, and radioactive—embed signals during training but involve trade-offs between effectiveness, security, and ethics. Notably, no single method fits all scenarios; future research should move beyond technical focus to consider fairness, accountability, and real-world impacts.
\end{tcolorbox}

\section{Maximally-Intrusive Protection}\label{sec:3}
Maximally-intrusive protection employs deliberate, aggressive modifications to datasets, aiming not only to embed ownership signals but also to actively disrupt unauthorized use. Unlike minimally-intrusive methods, these approaches apply substantial changes designed to enforce strict access control, often through reversible adversarial examples (RAEs).

Adversarial examples introduce carefully crafted perturbations to input data, causing machine learning models—especially deep neural networks—to make incorrect predictions, while leaving the inputs visually or semantically similar to the original \cite{szegedy2014intriguingpropertiesneuralnetworks,carlini2017towards}. However, traditional adversarial examples, while effective at deterring unauthorized model use, render the original data irrecoverable. To address this, Liu \textit{et al}.\cite{liu2023unauthorized} proposed the foundational RAE framework, integrating adversarial perturbations, reversible data hiding (RDH), and encryption, allowing authorized models to recover original images using a key while degrading unauthorized model performance. Building on this, Yin \textit{et al}.\cite{yin2019reversible} developed an RAE scheme based on reversible image transformation (RAE-RIT), which embeds indices of difference blocks between original and adversarial images for lossless recovery. To enhance generalization and robustness, Xue \textit{et al}.\cite{xue2023dataset} introduced random input transformations combined with feature space attacks, significantly improving the diversity and adaptability of RAE perturbations.

Further improvements in visual quality and embedding strategy have emerged. Yin \textit{et al}.\cite{yin2023reversible} proposed a YUV color space-based reversible attack that balances attack strength and visual fidelity while ensuring lossless recovery. Chen \textit{et al}.\cite{chen2024reversible} addressed scenarios involving locally visible adversarial perturbations, embedding recovery information into non-perturbed regions to enhance visual quality. Meanwhile, Cao \textit{et al}.\cite{cao2024reversible} reframed RAE generation as a deep steganography task, embedding self-embedding watermarks (W-RAE) to achieve both data access control and tamper resistance. Fang \textit{et al}.\cite{fang2023improving} contributed a probabilistic flip transformation method that improves RAE transferability under black-box conditions by optimizing the data hiding structure and leveraging data augmentation.

Black-box RAE generation presents additional challenges, as it requires crafting perturbations without internal model access. Xiong \textit{et al}.\cite{xiong2023black} proposed a black-box RAE (B-RAE) scheme that generates adversarial noise combined with RDH to maintain reversibility. Extending this, Huang \textit{et al}.\cite{huang2024black} developed I-RAE, using invertible neural networks and a Black-box Attack flow (BAFlow) to generate simple-distribution adversarial noise, reversibly embedded via a Perturbation Hiding Network (PHN) to achieve both strong attack capacity and high visual quality. Gao \textit{et al}.\cite{gao2021push} introduced the Transferable Attentive Attack (TAA), focusing on perturbing attention regions and features, with a triplet loss function that improves black-box attack success by steering source features away from their original class toward the target class.

To address unauthorized redistribution and copyright violations beyond mere model use, researchers have explored embedding visible watermarks and traceable signals. Chen \textit{et al}.\cite{chen2024rae} proposed RAE-VWP, which embeds visible watermark perturbations into salient image regions, combining privacy protection, copyright enforcement, and prevention of unauthorized distribution. Zhao \textit{et al}.\cite{zhao2022guided} developed the Guided Erasable Adversarial Attack (GEAA), embedding traceable watermarks into RAEs, enabling post-redistribution tracking and source identification.

In the context of creative industries, Shan \textit{et al}.\cite{shan2023glaze} introduced Glaze, a practical tool designed to protect artists’ unique styles from imitation by text-to-image (T2I) generative models. By adding minor perturbations to artworks, Glaze misleads generative models into learning incorrect stylistic features, safeguarding the artist’s visual signature without compromising the artwork’s perceptual quality.

\begin{tcolorbox}[title=Summary and Insight, colback=gray!10, colframe=black]
Maximally-intrusive methods, built on reversible adversarial examples, shift protection from passive signaling to active enforcement. They integrate perturbations, watermarking, and encryption to block unauthorized use, control access, and enable traceability. Yet they face persistent challenges: balancing robustness with visual quality, scaling to black-box settings, and managing computational cost.

Notably, these methods largely assume static adversaries and focus on technical barriers, leaving their long-term resilience and ethical impact underexplored. Future work should address not just stronger defenses but also how such aggressive protections reshape norms around data sharing, fairness, and trust in AI.
\end{tcolorbox}

\section{Challenge and Future Direction}\label{sec:challenge}
Dataset protection in machine learning faces several challenges that need to be addressed to ensure effective and scalable solutions.

One major challenge is the limited applicability of current methods. Most existing research has focused on image datasets, leaving other data types such as text, audio, and sensor data underexplored. As machine learning systems increasingly rely on diverse datasets, developing cross-domain protection techniques that can extend beyond image data is essential. This will be a critical direction for future research.

Another challenge is scalability. As datasets continue to grow in size, particularly with the rise of large-scale models, traditional protection methods, like watermarking, often struggle to scale efficiently. These methods can be computationally expensive when applied to large datasets, which is a significant barrier to widespread adoption. Future work should focus on scalable watermarking solutions that maintain effectiveness while minimizing computational overhead.

Privacy concerns also present a significant challenge. While watermarking and other protection techniques help ensure dataset ownership, they may raise issues regarding data privacy, especially with sensitive data. Developing privacy-preserving watermarking techniques that safeguard both ownership and privacy is a crucial research area. This will require innovation to balance dataset protection with the need to protect sensitive information.

Looking to the future, several directions emerge for further research. Cross-domain protection for non-image datasets is a priority, as the variety of data used in machine learning continues to grow. Specialized techniques for large-scale models, like dynamic watermarking or reversible data modifications, are needed to ensure dataset protection when data is generalized across large models. Blockchain technology also presents an opportunity to enhance transparency and accountability in dataset ownership. By integrating blockchain with traditional protection methods, we can create systems that offer more secure, decentralized, and transparent tracking of dataset usage.

Incorporating legal and ethical considerations into dataset protection methods is another key area for future work. Interdisciplinary collaboration will be essential to develop solutions that are not only technically effective but also legally sound and ethically responsible, ensuring privacy and fair use while maintaining dataset integrity.

\section{Conclusion}\label{sec:conclusion}
This survey explored a broad range of dataset copyright protection methods, from non-intrusive approaches like zero-watermarking and decision boundary analysis, to minimally-intrusive techniques such as reversible data hiding and dataset watermarking, and finally to maximally-intrusive strategies like reversible adversarial examples. Each method offers its own balance of strengths and limitations: non-intrusive methods preserve data untouched but may fall short against determined attackers; minimally-intrusive methods introduce lightweight signals but risk subtle trade-offs in robustness or ethics; maximally-intrusive methods actively block misuse but face practical concerns like visual distortion, transferability, and computational burden.

Most notably, current research is heavily focused on image datasets, with limited attention given to other types like text, audio, and sensor data. Additionally, the scalability of protection methods is a concern, especially as datasets grow in size and complexity with large-scale models.

Future research should focus on developing cross-domain protection solutions, improving scalability, and integrating privacy-preserving techniques to balance data protection with confidentiality. Furthermore, incorporating blockchain technology for dataset provenance tracking presents an exciting avenue to enhance dataset security and transparency.

\section*{Acknowledgment}
This work was funded by Key R\&D Program (Soft Science Project) of Shandong Province, China (2024RZB0406), the National Natural Science Foundation of China(No. 62402287, 62302266, 62232010, U23A20302, U24A20244), the Shandong Provincial Natural Science Foundation (No.ZR2024QF214), the Shandong Science Fund for Excellent Young Scholars (No.2023HWYQ-008), the project ZR2022ZD02 supported by Shandong Provincial Natural Science Foundation and the Guangdong Basic and Applied Basic Research Foundation (Grant No. 2025A1515010111).

%
%
%
\bibliographystyle{splncs04}
\bibliography{references}

\end{document}